\documentclass[10pt,conference]{IEEEtran}
\IEEEoverridecommandlockouts
\usepackage{cite}
\usepackage{amsmath,amssymb,amsfonts}
\usepackage{algorithmic}
\usepackage{graphicx}
\usepackage{textcomp}
\usepackage{xcolor}
\usepackage{float}

\def\BibTeX{{\rm B\kern-.05em{\sc i\kern-.025em b}\kern-.08em
    T\kern-.1667em\lower.7ex\hbox{E}\kern-.125emX}}
\begin{document}

\title{Quality Assurance of A GPT-based Sentiment Analysis System:\\ Adversarial Review Data Generation and Detection

}


\author{
Tinghui Ouyang$^{1}$, Hoang-Quoc Nguyen-Son$^{1}$,  Huy H. Nguyen$^{1}$, Isao Echizen$^{1}$, and Yoshiki Seo$^{2}$ \\
\small{\textit{$^{1}$National Institute of Informatics }} \\
\small{\textit{$^{2}$Digital Architecture Research Center, National Institute of Advanced Industrial Science and Technology, Japan}} \\
\small{E-mail: \{thouyang, nhhuy, iechizen\}@nii.ac.jp; y.seo@aist.go.jp}
}


\maketitle

\begin{abstract}
Large Language Models (LLMs) have been garnering significant attention of AI researchers, especially following the widespread popularity of ChatGPT. However, due to LLMs' intricate architecture and vast parameters, several concerns and challenges regarding their quality assurance require to be addressed. In this paper, a fine-tuned GPT-based sentiment analysis model is first constructed and studied as the reference in AI quality analysis. Then, the quality analysis related to data adequacy is implemented, including employing the content-based approach to generate reasonable adversarial review comments as the wrongly-annotated data, and developing surprise adequacy (SA)-based techniques to detect these abnormal data. Experiments based on Amazon.com review data and a fine-tuned GPT model were implemented. Results were thoroughly discussed from the perspective of AI quality assurance to present the quality analysis of an LLM model on generated adversarial textual data and the effectiveness of using SA on anomaly detection in data quality assurance.
\end{abstract}


\section{Introduction}
\label{sec:intro}

ChatGPT, developed and released by OpenAI \cite{b1}, has emerged as the hottest ``super-star'' in the field of artificial intelligence (AI) currently. It is primarily based on LLMs and a variant of the Generative Pre-trained Transformers (GPT) \cite{b2}. Due to its friendly interface and tremendous in-context learning capability, ChatGPT has achieved great success in various natural language process (NLP) tasks, like question answering, chatting, grammar checking, and so on. For example, ChatGPT was utilized as a translator in \cite{b3}. In \cite{b4}, it was also used to check the grammar errors in writing with a comparative performance with commercial products. Moreover, another success of ChatGPT usage is in the education field \cite{b05}-\cite{b6}, like exam testing or plagiarism. 

However, as a kind of LLM-based product, its vast parameters and complex architecture bring several challenges regarding AI quality management (AIQM) in the usage of ChatGPT, such as concerns related to security, privacy protection, attacks, and the risks of misuse. For instance, using ChatGPT for plagiarism is a severe issue of scientific integrity. Therefore, adhering to the standard guideline of AI quality assurance \cite{b7}, it becomes imperative to thoroughly investigate the quality issues associated with ChatGPT and other LLMs based on the GPT framework. Currently,  the quality issues of ChatGPT and LLMs have attracted many scholars' attention to research. For example, the coverage of ChatGPT was tested on high-resource and low-resource languages, respectively \cite{b3}. The ability of ChatGPT on code and mathematics was also investigated \cite{b8}. Moreover, by taking ChatGPT as an AI product, its qualities of reliability and robustness were studied in \cite{b9}-\cite{b10}.

In this paper, a specific NLP topic focusing on sentiment analysis is selected as a reference for the AIQM study on LLMs. Sentiment analysis is the simplest NLP task, which aims to classify a piece of text into different sentiment categories, such as positive, negative, and neutral. Companies will use these comments to evaluate customers' satisfaction with products/services, guide manufacturing and production, and predict market movements. However, as reported in \cite{b11}-\cite{b13}, it is super easy for customers to generate fake comments in the real world when reviewing products on e-commercial platforms, especially with the help of ChatGPT. If these fake comments are consistent with the rating customers gave, it is not a big issue. If the comments are randomly generated with mismatched rating levels, it will bring a high risk of economic loss to product manufacturers. A lot of such cases were reported in \cite{b13b}. For example, in the context of an app store, if the rating level of a particular app does not align with the review comment, it can have a detrimental impact on its future downloads. This discrepancy can lead to a loss in profits, particularly for upcoming apps with fewer comments. Furthermore, the prevalence of dishonest reviews and unfair ratings has emerged as a significant problem in practical commercial sales \cite{b14}. Therefore, it is seen that detecting fake comments with unjustifiable labels has become an intriguing area of research in AIQM, especially data quality assurance. 

In AIQM, data quality analysis suggests that one straightforward approach to detect these wrongly-annotated review data is through examining data distribution \cite{b18}. However, textual data, being non-numeric and having complex properties, present challenges in distribution analysis. This is also the reason why NLP studies usually use some techniques to transform text into numeric data first, like TF-IDF, bag-of-words, word2vec, or other tokenization processes \cite{b15}-\cite{b17}. Moreover, the transformed numerical vectors may have sparsity and high dimensionality, so it is not directly feasible for data distribution analysis. With the above considerations, the other type of method to detect these wrongly-annotated data is through misclassification, which assumes the given sentiment analysis system is confident to make correct decisions. For example, a deep learning model based on a recurrent neural network (RNN) with a gate recurrent unit was constructed for sentiment analysis, and instances of incorrect decisions are attributed to the wrongly-annotated data \cite{b19}. In \cite{b20}, three systems using different ML/DL algorithms were developed to detect the wrongly-annotated data in Android app reviews. Moreover, review data were categorized into unfair sentiment comments based on the model's misclassification in \cite{b21}. However, the related question is how to guarantee the trained AI model is truly confident in sentiment analysis. If wrongly-annotated data exists in the training dataset without notification, the practical training process and decision patterns of AI will also be influenced by those mismatched data. Then, it is hard to guarantee the confidence and quality assurance of the trained AI model.

Based on the statement and discussion above, this paper proposes to investigate the wrongly-annotated data and their influence on a given LLM model for AIQM. First, we orient to the sentiment analysis task and build an LLM model. Aiming at the reference study, a basic GPT model released by OpenAI is fine-tuned on the Amazon.com review data which is publicly available and widely used as the reference in sentiment analysis. Then, we implement some possible AI quality assurance studies based on the constructed LLM model, e.g., correctness, robustness and data adequacy analysis. One study is to acquire the wrongly-annotated review comments. This paper proposes to apply the content-based method to generate adversarial textual data, which is required to have similar data qualities with the original ones but leads to incorrect decisions. The other study is to develop methods for detecting these abnormal data. This paper develops a new metric based on surprise adequacy (SA) \cite{b22}, and evaluates its performance on abnormal data detection and data quality assurance.

The rest of this paper is organized as follows. 
Section~\ref{sec:definition} presents the general definition of adversarial samples in deep neural networks (DNN). 
Section~\ref{sec:generation} introduces the methods for generating adversarial textual data, especially the wrongly-annotated review comments in this paper. 
Section~\ref{sec:detection} proposed advanced techniques to detect those adversarial data for quality analysis on data adequacy. 
Experiments and numerical comparisons on GPT-based sentiment analysis are implemented in Section~\ref{sec:experiments}. 
Finally, the results and conclusions are summarized in the Section~\ref{sec:conclusions}.

\section{Definition of DNN's adversarial data}
\label{sec:definition}
According to the above description, it is easily seen that the wrongly-annotated review comments share a similar concept with adversarial data. In AI quality assurance \cite{b24}-\cite{b25}, adversarial data have been widely studied as special data samples fooling a neural network. Typically, in the context of recognition and classification, the fooling behaviors can result in misclassification, so they tightly correlate to data adequacy and data security in AIQM.

To formulate the adversarial examples in mathematics, we can first assume a deep neural network (DNN) for classification or recognition as $F(\cdot): R^d\rightarrow[0,1]^m$. This model aims to map a $d$-dimensional input $x$ to the probability vector $F(x)$ with a dimensionality of $m$, where $m$ is the number of output labels. The input can be numeric data, image data, or textual data based on the studied scenarios, and the output is the probability distribution of all possible classification labels. Let the output of the last layer of DNN be denoted as $Z(\cdot)$, then the label probability of DNN is usually calculated by the softmax activation function of $Z$, expressed below.
\begin{equation}
F(x)=softmax(Z(x))
\end{equation}

The predicted class label of $x$ can be determined by the maximum probability in $F(x)$, namely
\begin{equation}
C(x) ={\operatorname{argmax}}\ F(x)
\end{equation}

Generally, to train this classification model, supervised learning is implemented in the training process along with a set of training samples consisting of $(x,y)$ where $y$ is the target class label.

Then, based on the definition of adversarial samples, the following equation can formulate the adversarial samples.
\begin{equation}
\left\{\begin{array}{l}
\hat{x}=x+\delta \\
s.t.\ C(\hat{x}) \neq C(x)\\
\end{array}\right.
\end{equation}

This equation illustrates that adversarial samples are actually produced by original samples with additional small perturbation $\delta$. This perturbation is usually thought imperceptible to human beings. However, they can lead the trained DNN $F(\cdot)$ to an incorrect label.

\section{Adversarial textual comments generation}
\label{sec:generation}
Based on the definition and formulation of adversarial samples, three categories of adversarial samples generation methods can be summarized by considering the information of trained DNN model \cite{b26}, such as zero-knowledge, limited- knowledge and perfect-knowledge methods. The first type aims to generate adversarial samples on an unsecured model $F(\cdot)$ and ignore whether an attack detector exists or not. The second one is trained to generate adversarial samples fooling the model $F(\cdot)$ but with limited knowledge. For example, it is known that there exists a detector securing the model $F(\cdot)$ against a given attacker, but no details about the detector. Unlike the second type, the final one has the knowledge of both the detector's existence and its architecture and parameters. Therefore, it can use ``perfect knowledge'' to generate adversarial samples to bypass the given detector.

Currently, in adversarial data generation study, the popular way is to use the gradient to optimize the perturbation $\delta$, e.g., using $L_{0-2}$ norm in objective functions. Some famous methods, like FGSM/BIM/PGD and DeepFool \cite{b27}-\cite{bdf}, have been widely applied in image-based recognition systems. For textual data, a gradient-based method is also applicable \cite{b30}, though gradient-based methods are commonly implemented based on numeric data. While NLP studies usually involve tokenization and reverse-tokenization processes to realize the transformation between text and numeric data, it is not completely feasible to guarantee the same semantic information of original data in gradient-based adversarial text generation. That means the perturbation generated by gradient-based methods may not be imperceptible to humans. With this consideration, this paper proposes to adopt another type of adversarial data generation method, namely the content-based method \cite{b31}. 
This method leverages the input content and adds restrained perturbations required to provide semantically consistent with simulating real-world context. For example, concerning image data collected from cameras, it is possible to have some black spots in an image due to a dirty camera lens. Therefore, attackers can easily add black spots to images to generate adversarial samples instead of using gradient-based methods. Some examples of this kind of attack were reported in literature \cite{b32}-\cite{b33}.

Considering the target of this paper is to study the AIQM in an LLM-based model, especially the data quality assurance in sentiment analysis, the targeted objective is textual review comments from the real world. Therefore, if using the content-based method for adversarial data generation, it is possible to have some typos in the text due to spelling errors or have different contractions on expression, e.g., using "hasn’t" to replace "has not", as presented below.

\begin{figure}[H]
\centering
\includegraphics[width=0.5\textwidth]{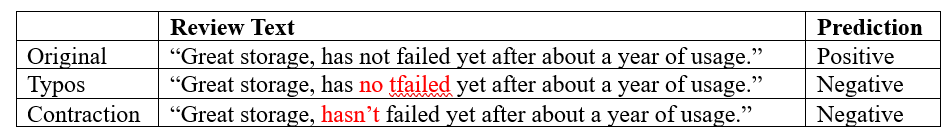}
\end{figure}
\begin{figure}[H]
\centering
\includegraphics[width=0.5\textwidth]{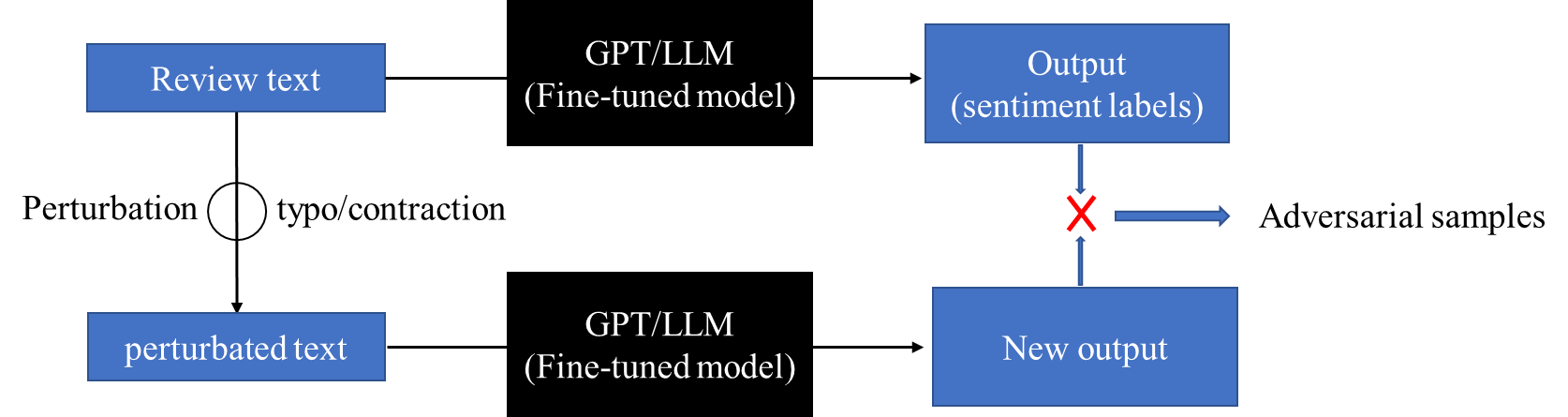}
\caption{Framework of adversarial data generation in GPT-based model}
\label{fig2}
\end{figure}

These cases reasonably exist in the real world with high possibility, and they can surely preserve the semantic information of original data. Therefore, we can use the content-based method to generate some adversarial textual data fooling a given AI system, e.g., the GPT-based sentiment analysis model. To make use of the content-based method in adversarial text generation, this paper proposed a framework as Fig. \ref{fig2}
According to the proposed framework, it is seen that there are mainly two parts:

(1)	Fine-tuning a GPT-based model for sentiment analysis

 In this paper, we first need to obtain a sentiment analysis model as the reference in AIQM. Therefore, to evaluate the quality of LLMs, we choose to fine-tune a model based on the GPT-3 released by OpenAI. Here, the specific scenarios can be chosen as the sentiment analysis on Amazon.com review comments, so the public Amazon review dataset is chosen as the training dataset.

(2)	Generating perturbed textual data

To add perturbations to text according to the content-based method, we can use the useful NLP package "CHECKLIST" \cite{b34}, which is easily used to add a different number of typos and contraction operations to textual data. Then, we can feed these perturbed data to the fine-tuned GPT model and compare the classification results. If a mismatched classification label is outputted, we can collect it as an adversarial candidate for the quality study in the next steps.

\section{Adversarial review comments detection}
\label{sec:detection}
After generating adversarial review comments, we can detect these data for data adequacy analysis. Considering that ChatGPT API or GPT model provided by OpenAI seems like a black box, this paper proposed using surprise adequacy as the tool to develop adversarial data detection methods. Surprise adequacy (SA) was initially proposed \cite{b22} to measure the novelty between testing data and training data, and it was also verified helpful in AI quality assurance \cite{b35} as well as in NLP study \cite{b18}. The detailed processes are described as follows to calculate the value of SA.

1) Neuron activation status

\begin{figure*}[htbp]
\centering
\includegraphics[width=0.85\textwidth]{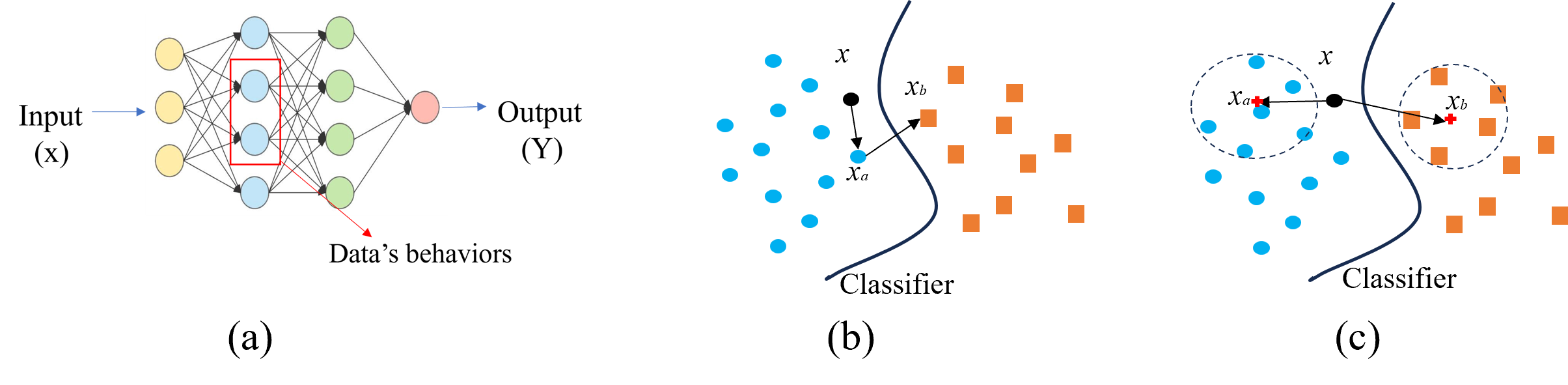}
\caption{Calculation of distance-based surprise adequacy. (a) activation trace; (b) original DSA; (c) improved DSA.}
\label{f2}
\end{figure*}

It is known that the easy way of comparing data differences is through their numeric values. According to the definition of adversarial samples, we see a significant difference in adversarial data happening on their behavior with respect to a given AI model. In SA calculation, the neuron activation status is usually used to describe the behavior of data, as shown in Fig. \ref{f2}(a). Assuming an AI system consisting of a set of neurons $N=\{n_1, n_2,\cdots\}$, then the output value or status of a given neuron $n\in N$ with respect to a testing data $x$ can be directly regarded as the data's behavior, denoted as $\varphi(x)$. Then, for a sequence of neurons in an organized architecture, like a layer or graph, the neuron activation status of the testing data is meaningful in AI's quality analysis. Assuming a subset $N_S\subseteq N$, its neuron behaviour on $x$ can be easily expressed as
\begin{equation}
\Phi_{N_S}(x)=[ \varphi_{n_1}(x), \varphi_{n_2}(x),\cdots, \varphi_{n_k}(x)]^T, n_k\in N_S
\end{equation}

Then, based on the behaviors on a given subset $N_S$, the difference between testing data and training data could be calculated, namely the surprise studied in SA. Therefore, SA can be expressed in the following form.

\begin{equation}
SA=sp(\Phi_{N_S}(x_t), \Phi_{N_S}(D_T))
\end{equation}
where, $sp(\cdot)$ is the surprise function. $x_t$ and $D_T$ represent the testing and training data, respectively. For the consideration of simplifying calculation, the subset $N_S$ can be chosen as neurons of a given layer. Then, it is easily understood that SA is namely to compare the AI model’s reaction to testing data and that of training data, so it is useful in software testing and quality assurance study.  

2) Distance-based SA (DSA)

According to the above definition, there are two types of SA defined in \cite{b22}, such as likelihood-based SA (LSA) and distance-based SA (DSA). While considering DSA is more suitable in classification problems, e.g., sentiment classification, so here DSA is introduced in this paper. The definition of DSA is expressed in (\ref{eq1}) and shown in Figure \ref{f2}(b).
\begin{equation}
DSA(x)=\frac{dist_a}{dist_b}
\label{eq1}
\end{equation}

It is seen that DSA is defined as the ratio between two distances, namely meaning that the function $sp(\cdot)$ is selected as a distance function. In \cite{b22}, these two distances are defined as the following equations

\begin{equation}
\left\{\begin{array}{l}
dist_a=\Vert \Phi (x_t)-\Phi (x_a)\Vert \\
dist_b=\|\Phi (x_a )-\Phi (x_b)\| 
\end{array}\right.
\label{eq2}
\end{equation}
where, $x_t$ is a testing data point in the class $y_t$. $x_a\in D_T$ is the nearest data point of $x_t$ in the same class $y_t$ of training dataset, defined as 
\begin{equation}
x_a=\underset{C({x_i})=y_t}{\operatorname{argmin}}\Vert \Phi (x_t)-\Phi (x_i)\Vert                  
\label{eq8}
\end{equation}

In (\ref{eq2}), $x_b$ is the nearest data point of $x_a$ in another classes of $D_T$, expressed as
\begin{equation}
x_b=\underset{C({x_i})\in \{C-y_t\}}{\operatorname{argmin}}\|\Phi (x_a )-\Phi (x_i)\|     \label{eq9}          
\end{equation}

\noindent where, $\{C-c_t\}$ represents the set of all classes different with $c_t$. According to the above definitions, it is seen that the DSA value will be high when a large value of numerator and a small value of denominator in (\ref{eq1}). That implies the testing data $x_t$ is distant to its labelled class $c_t$ and close to data of other classes. In this case, it is reasonable to say it's surprised to normal data of its belonging class, so having high risk of being anomaly. Based on this idea, DSA was verified useful to describe anomaly data’s behaviours respect to a given DL model, and to detect corner case data \cite{b36}.

3) DSA modification

While from the original definition of DSA, we see that SA is actually to compare the surprise of $x_a$ (the nearest training data point of $x_t$ in the same class) with that of other classes. Therefore, it can be simplified as the comparison between $x_t$’s novelty in its labeled class and its novelty in other classes. This definition is helpful to describe the surprise of testing data to the training dataset. However, there are some shortages in the calculation of $dist_b$ found. For example, if a data point belongs to corner case data \cite{b36}, its novelty with respect to all classes seems more critical. Moreover, considering the distances of (\ref{eq8})-(\ref{eq9}) are processing pair-wise rare data points, if the testing data $x_t$ is an extreme point and happens to have a neighbor $x_a$ very close to it, $x\approx x_a$, then the calculation of DSA will be a low value $dist_a\rightarrow0$ which will mislead the surprise measurement, especially on abnormal data detection. Therefore, with the above consideration, this paper proposed to improve the calculation of DSA, especially modifying the measure of $dist_a$ and $dist_b$ to improve the generability of DSA, as expressed below.

\begin{equation}
\left\{\begin{array}{l}
dist_a=\left\|\Phi\left(x_t\right)-\Phi\left(x_a\right)\right\| \\
dist_b=\left\|\Phi\left(x_t\right)-\Phi\left(x_b\right)\right\|
\end{array}\right.
\end{equation}
where the calculation of $x_a$ and $x_b$ will be modified as below
\begin{equation}
\begin{aligned}
x_a=\frac{1}{|X_a|} \sum_{x_i \in X_a}{x_i},&\ X_a=\{x_i|\ C({x_i})=y_t \ \& \ {x_i\in Nb(x_t)}\} \\
x_b=\frac{1}{|X_b|} \sum_{x_i \in X_b}{x_i},&\ X_b=\{x_i|\ C({x_i})\in \{C-y_t\} \ \& \ {x_i\in Nb(x_t)}\}  
\end{aligned}
\label{eq11}
\end{equation}

\noindent where $Nb(x_t)$ is defined as a function of finding the neighborhood of the testing data $x_t$ with a given condition; $|X|$ means the cardinality of the set $ X$. Here, the modified definition of $x_a$ is to find the center of $x_t$'s neighborhood in the same class $y_t$. Meanwhile, $x_b$ is the center of $x_t$'s neighborhood in the other class, as shown in Fig. \ref{f2}(c).
With the help of the above modification, some drawbacks of the original DSA can be eliminated. Moreover, in terms of some special cases, more variants of DSAs can also be developed based on the description in Figure \ref{f2}(c), e.g., using global or local neighborhood in the calculation of $ Nb(\cdot)$. Then, using these DSAs, we can realize the abnormal data detection in AI quality assurance, like adversarial data detection for data adequacy study.

\section{Experiments}
\label{sec:experiments}
To study the AI quality issues of LLM, this paper chooses the simplest task of sentiment analysis as an example. In accordance with the framework in Fig. 1, the AI model studied in this paper is a GPT-3 model from OpenAI and fine-tuned on the well-known Amazon.com review dataset. Then, we can generate some adversarial review comments and evaluate the performance of GPT-based LLM on data adequacy analysis. Aiming at the mentioned task, this paper mainly addresses three questions listed as below.

1) Correctness study of the fine-tuned GPT model

2) Adversarial textual data generation and robustness analysis

3) Adversarial data detection for data quality assurance

\subsection{Correctness analysis}

Currently, it seems that ChatGPT can help humans to do any NLP jobs under its popularity. With the help of the foundation model, OpenAI also provides developers with commercial ChatGPT API for research. Therefore, it is easy to use ChatGPT API directly to realize the sentiment analysis on Amazon.com review data. However, as a general model, the ChatGPT API has no defined boundary for specific sentiment categories. To achieve accurate classification on a particular task, OpenAI also provides the other option to developers, namely using their existing GPT model for fine-tuning on a given task. In this paper, both of these two options are considered in modeling the sentiment analysis task on Amazon.com review data. Their performances are shown below.

\begin{figure}[H]
\centering
\includegraphics[width=0.5\textwidth]{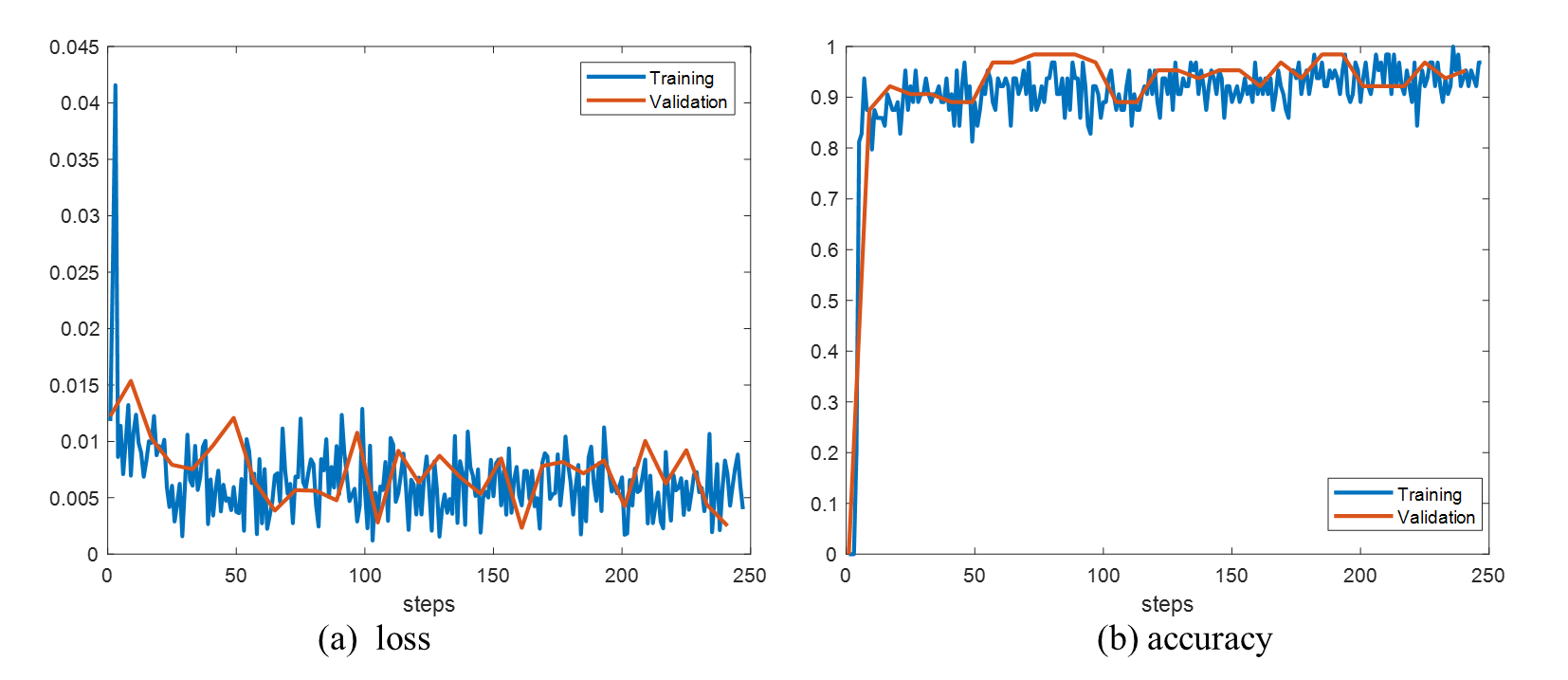}
\caption{Performance of the fine-tuned GPT model on sentiment analysis}
\label{f3}
\end{figure}

Fig. \ref{f3} shows the performance of the fine-tuned GPT model on both the training and testing process. It is seen that the GPT model can obtain good performance on sentiment analysis after fine-tuning the Amazon.com review dataset, seeing the decreasing loss and increasing accuracy in Fig. \ref{f3}(a)-(b), respectively. Moreover, to quantitatively investigate the performance of studied models, Table \ref{t1} presents the testing accuracy of both using ChatGPT API and fine-tuned GPT model in sentiment analysis.

\begin{table}[H]
\centering
\caption{Accuracy of sentiment analysis models}
\begin{tabular}{ccc}
\hline
 & ChatGPT API & Fine-tuned GPT\\
 \hline
Accuracy & 84.54\% &94.91\%\\
\hline
\end{tabular}\label{t1}
\end{table}
Through the comparative analysis, it is seen that the fine-tuned model has great improvement in accuracy compared with the method using ChatGPT API directly for sentiment analysis. Therefore, it illustrates the necessity of fine-tuning when making use of LLM to develop some specific NLP tasks in real engineering. Moreover, it is reasonable for us to take the fine-tuned GPT model as the reference in AI quality analysis in sentiment analysis.
\subsection{Adversarial data generation and robustness analysis}
Considering that wrongly-annotated data is a realistic data quality problem in AIQM, these data usually harm modeling and detection. However, the knowledge of these data is yet to be discovered by developers, and it usually requires enormous manual efforts for annotation. For example, regarding the Amazon.com review data, the raw data are obtained from realistic customer reviews. Sometimes customers make these wrongly-annotated data by mistakes, so it is unaware to pick them up in data preparation. The other issue is about labeling. This process is subjective since the sentiment labels of review comments are usually determined by the rating levels customers give, e.g., the rating stars. With the above considerations, this paper proposes to generate some adversarial review comments instead of the realistic wrongly-annotated data for data quality analysis in Section~\ref{sec:generation}. Then, aiming at the given reference AI model, namely the fine-tuned GPT-based sentiment analysis model, this paper adopts the content-based method to generate adversarial review comments. In this paper, using the ``CHECKLIST'' package, different numbers of typos are added to the original review comments to generate adversarial samples. The results are shown below.
\begin{table}[H]
\centering
\caption{attack success rate}
\begin{tabular}{cccccc}
\hline
	&1 typo	&2 typos	&3 typos	&4 typos	&5 typos\\
\hline
ASR&0.0702	&0.0814	&0.0804	&0.0732	&0.1068\\
\hline
\end{tabular}\label{t2}
\end{table}
Table \ref{t2} shows the successful attack rate (ASR) under different numbers of typos in perturbation. It is seen that the GPT-based model is generally robust again the content-based attack, like the typo perturbation, since the attacked samples have a proportion of around 10\% in Table \ref{t2}. To further study these perturbed data's properties, the text length of successfully-attacked review comments is summarized in the following figure.
\begin{figure*}
\centering
\includegraphics[scale=0.75]{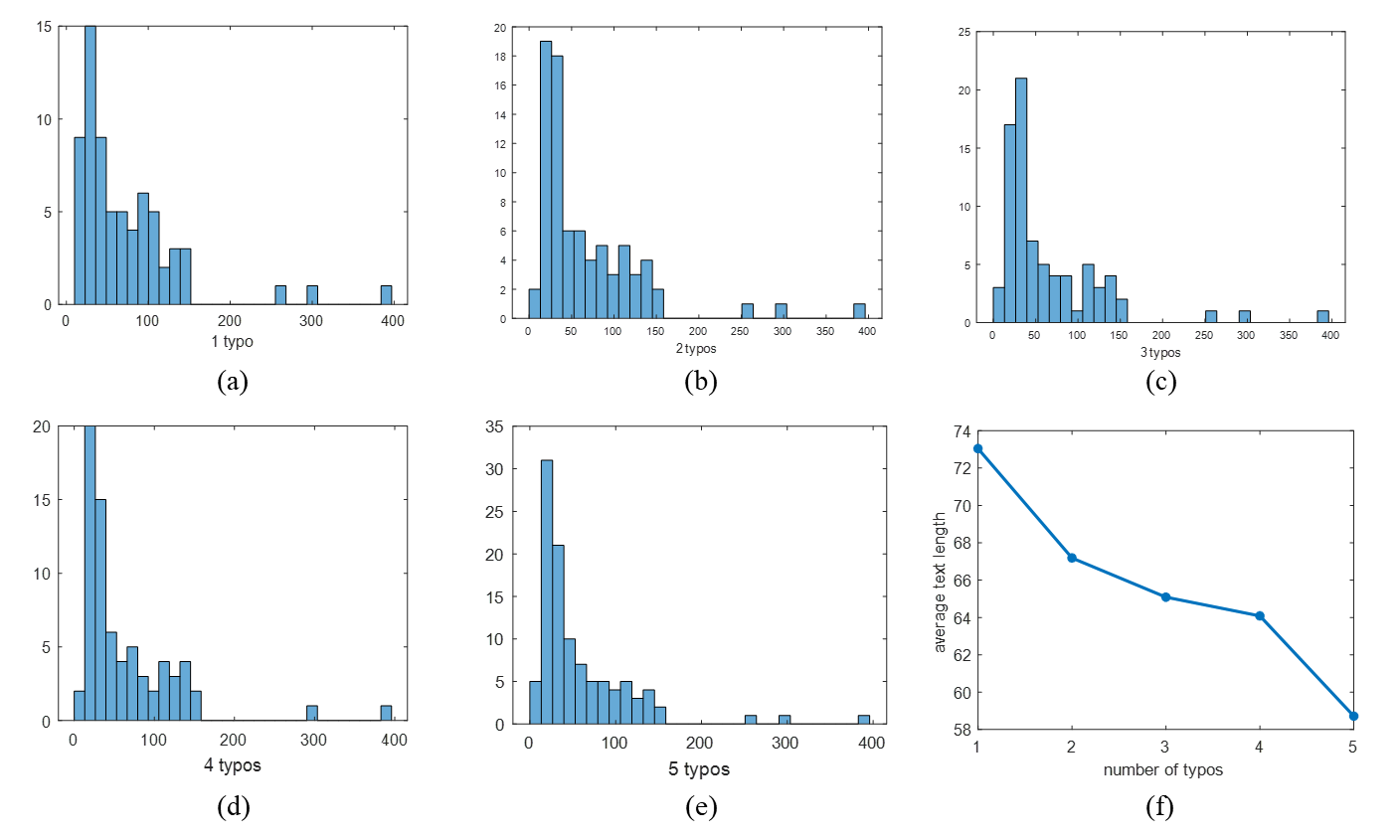}
\caption{The statistics of text length under successful attacks}
\label{f4}
\end{figure*}
From the above figures, it is seen that successful attacks usually happen on data with short text length, implying long review comments are robust to typo perturbation. Moreover, Fig. \ref{f4}(f) shows the average text length of successful attacks under different numbers of typos. It is seen, as the number of typos increases in perturbation, the average text length of attached review comments decreases. It implies that more short review comments are attacked when more typos are added. It further demonstrates the conclusion that short review comments are less robust than long review comments, which is reasonable with human beings' knowledge in the real world.

\subsection{Adversarial data detection for data quality assurance}
\begin{figure*}
\centerline{\includegraphics[width=1\textwidth]{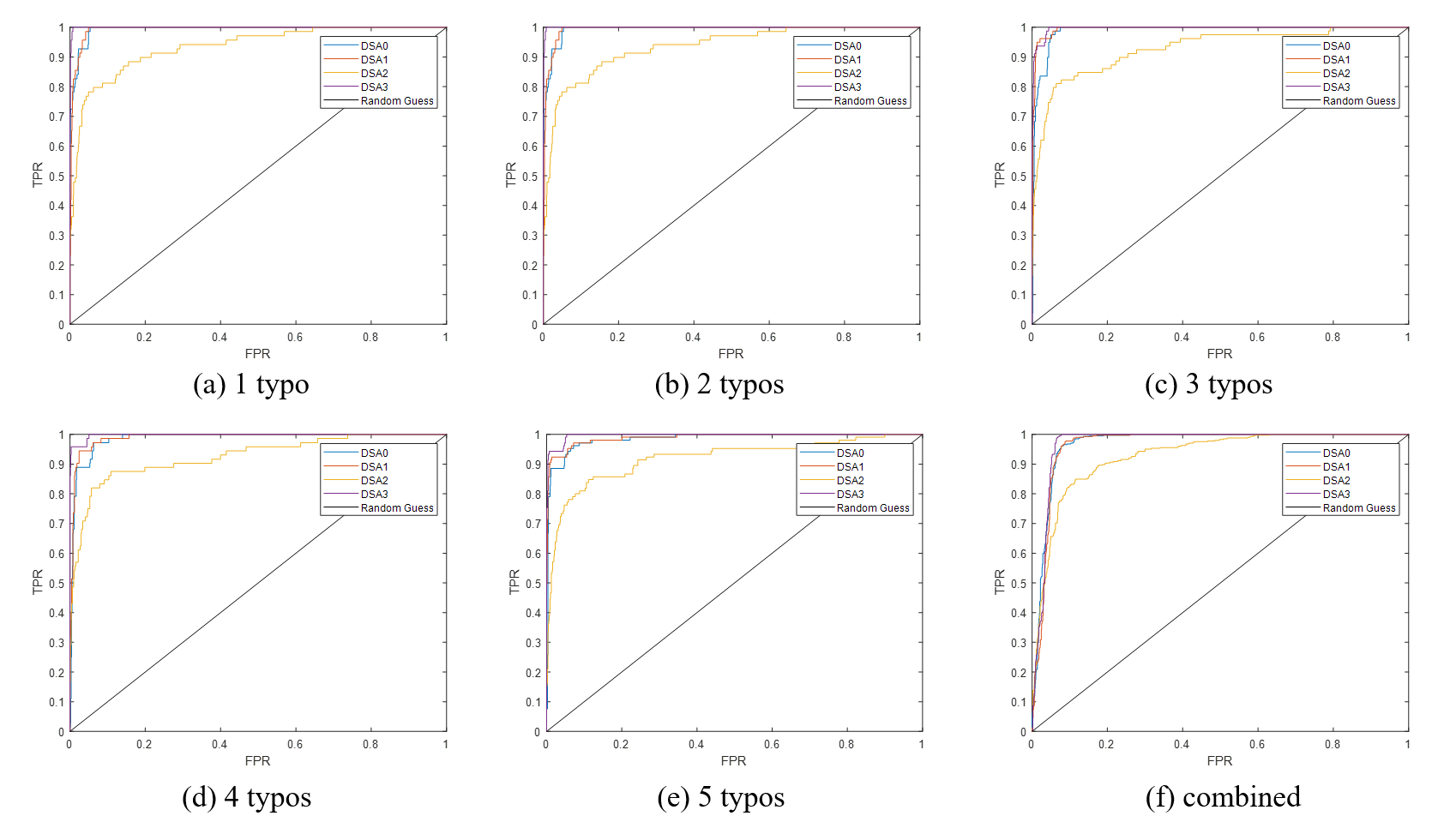}}
\caption{Performance of different DSA metrics on adversarial text detection}
\label{f5}
\end{figure*}

Moreover, considering the importance of data adequacy quality analysis, this paper also proposed advanced anomaly detection methods oriented to the generated adversarial review comments. Section~\ref{sec:detection} presents the original and modified distance-based surprise adequacy (DSA) metrics for anomaly detection. In this paper, to evaluate the AI model's quality thoroughly, several variants of DSAs are adopted in the experiment for comparative study. For example, the original DSA is selected as the reference in adversarial textual data detection, denoted as DSA0. The modified DSA can be represented as DSA1, which is set up with the simplest parameters, namely using the nearest point in the calculation $x_a$ and $x_b$ in (\ref{eq11}). Moreover, considering class centers are common helpful global descriptors \cite{b37}, we can also modify the calculation in (\ref{eq11}) with the whole class of data as $X_a$ and $X_b$, and denote the DSA variant as DSA2. For comparison, a version of using a local descriptor, like utilizing a neighborhood consisting of 10-nearest points as $X_a$ and $X_b$ in (\ref{eq11}), is also applied in this paper, denoted as DSA3. Then, by extracting the hidden layer's output as data's behaviors, these four DSAs are calculated. Their performance on adversarial review comments detection is shown via ROC curves, as shown in Fig. \ref{f5}.

To quantitative analyse their performance, numerical results are computed via AUC-ROC values are shown in the following table.
\begin{table}
\centering
\caption{Numerical results of adversarial text detection}
\begin{tabular}{ccccccc}
\hline
	&1 typo	&2 typos	&3 typos	&4 typos	&5 typos	&combined\\
\hline
DSA0	&0.9931	&0.9939	&0.9879	&0.9857	&0.9852	&0.9683\\
\hline
DSA1	&0.9927	&0.9973	&0.9945	&0.9894	&0.9893	&0.9653\\
\hline
DSA2	&0.9345	&0.9311	&0.9270	&0.9256	&0.9188&	0.9277\\
\hline
DSA3	&\textbf{0.9994}	&\textbf{0.9980}&\textbf{0.9964}	&\textbf{0.9978}	&\textbf{0.9966}	&\textbf{0.9707}\\
\hline
\end{tabular}
\label{tb3}
\end{table}
In Fig. \ref{f5} and Table \ref{tb3}, four DSA variants are measured based on adversarial data generated with different numbers of typos. Moreover, a comprehensive study combining these generated adversarial review comments together is also implemented in this section. It is seen from these results that DSA3 outperforms the other DSAs, and DSA2 performs the worst via comparison. These results illustrate that a local descriptor in DSA measurement is better than a global one. Moreover, the modified DSA can outperform the original one on adversarial data detection via optimized parameters. Therefore, in the data adequacy study on GPT-based sentiment analysis, we could select optimized DSA for abnormal data detection and extract the possible wrongly-annotated review comments for the GPT-based model's correctness and robustness analysis.

\section{Conclusions}
\label{sec:conclusions}
In this paper, to investigate the AI quality assurance issue of LLMs, this paper proposes to fine-tune a GPT model as the reference. Orienting to classify the sentiment categories of Amazon.com review comments, a GPT-3 model provided by OpenAI is fine-tuned. Then, reasonable adversarial data based on the content-based method are generated for data adequacy analysis. These adversarial texts well preserved the semantic information of the original text by using typo perturbations. The reference GPT model is demonstrated to be robust in these adversarial attacks. Moreover, using SA to describe the data's novelty and surprise on the model's behaviors, some variants of DSA were developed in this paper and used to detect the generated adversarial review comments. Finally, numerical results show that modified DSA with optimized local descriptors can achieve good performance in detecting adversarial data and possibly detect wrongly-annotated review comments in sentiment analysis. Summarizing all of these results, it is seen that the fine-tuned GPT model performs well on sentiment analysis, and it is robust to natural adversarial review comments generated by the content-based method. Moreover, the proposed DSAs in this paper are feasible and effective in detecting the adversarial data in the AIQM study of GPT-based models and especially useful for the data quality assurance quality. Furthermore, based on the results of this paper, it is possible to study some extra AI qualities with the detected adversarial textual data in the future. 

\section*{Acknowledgement}
\label{sec:acknowledgement}
This research is supported by the project 'JPNP20006', commissioned by the New Energy and Industrial Technology Development Organization (NEDO), and partly 
supported by JSPS Grant-in-Aid for Early-Career Scientists (Grant Number 22K17961)


\begin{thebibliography}{00}
\bibitem{b1} OpenAI. https://chat.openai.com.chat, 2023.
\bibitem{b2} Long Ouyang, Jeff Wu, Xu Jiang, Diogo Almeida, Carroll L Wainwright, Pamela Mishkin, Chong Zhang, Sandhini Agarwal, Katarina Slama, Alex Ray, et al. Training language models to follow
instructions with human feedback. arXiv preprint arXiv:2203.02155, 2022.
\bibitem{b3} Jiao, W. X., W. X. Wang, J. T. Huang, Xing Wang, and Z. P. Tu. Is ChatGPT a good translator? Yes with GPT-4 as the engine, 2023. arXiv preprint arXiv:2301.08745.

\bibitem{b4} Wu, Haoran, Wenxuan Wang, Yuxuan Wan, Wenxiang Jiao, and Michael Lyu. ChatGPT or Grammarly? Evaluating ChatGPT on Grammatical Error Correction Benchmark, 2023. arXiv preprint arXiv:2303.13648.

\bibitem{b05} Jalil, Sajed, Suzzana Rafi, Thomas D. LaToza, Kevin Moran, and Wing Lam. Chatgpt and software testing education: Promises \& perils, 2023. arXiv preprint arXiv:2302.03287.
\bibitem{b6} Fitria, T. N.  Artificial intelligence (AI) technology in OpenAI ChatGPT application: A review of ChatGPT in writing English essay. In ELT Forum: Journal of English Language Teaching 12(1), pp. 44-58, 2023.
\bibitem{b7} Machine Learning Quality Management Guideline (https://www.digiarc.aist.go.jp/en/publication/aiqm/)
\bibitem{b8} Frieder, Simon, Luca Pinchetti, Ryan-Rhys Griffiths, Tommaso Salvatori, Thomas Lukasiewicz, Philipp Christian Petersen, Alexis Chevalier, and Julius Berner. Mathematical capabilities of ChatGPT, 2023. arXiv preprint arXiv:2301.13867.
\bibitem{b9} Reiss, Michael V. Testing the Reliability of ChatGPT for Text Annotation and Classification: A Cautionary Remark, 2023. arXiv preprint arXiv:2304.11085 .

\bibitem{b10} Wang, Jindong, Xixu Hu, Wenxin Hou, Hao Chen, Runkai Zheng, Yidong Wang, Linyi Yang et al. On the robustness of chatgpt: An adversarial and out-of-distribution perspective, 2023. arXiv preprint arXiv:2302.12095.

\bibitem{b11} Mitrović, Sandra, Davide Andreoletti, and Omran Ayoub. Chatgpt or human? detect and explain. explaining decisions of machine learning model for detecting short chatgpt-generated text, 2023. arXiv preprint arXiv:2301.13852.
\bibitem{b12} Dai, Haixing, Zhengliang Liu, Wenxiong Liao, Xiaoke Huang, Zihao Wu, Lin Zhao, Wei Liu et al. Chataug: Leveraging chatgpt for text data augmentation, 2023. arXiv preprint arXiv:2302.13007.
\bibitem{b13} Adelani, David Ifeoluwa, Haotian Mai, Fuming Fang, Huy H. Nguyen, Junichi Yamagishi, and Isao Echizen. Generating sentiment-preserving fake online reviews using neural language models and their human-and machine-based detection. In Advanced Information Networking and Applications: Proceedings of the 34th International Conference on Advanced Information Networking and Applications (AINA-2020), pp. 1341-1354. Springer International Publishing, 2020.

\bibitem{b13b} Kushal Dave, Steve Lawrence, and David M. Pennock. Mining the Peanut Gallery: Opinion Extraction and Semantic Classification of Product Reviews. In Proceedings of the 12th International Conference on World Wide Web (WWW ’03). ACM, New York, NY, USA, 519–528. 2003. https://doi.org/10.1145/775152.775226
\bibitem{b14} Diekmann, A., B. Jann, W. Przepiorka and S. Wehrli, Reputation formation and the evolution of cooperation in anonymous online markets. Am. Sociol. Rev., 79: 65-85, 2014.
\bibitem{b18} Ouyang, T., Seo, Y., and Oiwa, Y.  Quality assurance study with mismatched data in sentiment analysis. In 2022 29th Asia-Pacific Software Engineering Conference (APSEC) (pp. 442-446). IEEE. 2022, December.

\bibitem{b15} Qaiser, Shahzad, and Ramsha Ali. Text mining: use of TF-IDF to examine the relevance of words to documents. International Journal of Computer Applications 181, no. 1: 25-29, 2018.
\bibitem{b16} Zhang, Yin, Rong Jin, and Zhi-Hua Zhou. Understanding bag-of-words model: a statistical framework. International journal of machine learning and cybernetics 1, no. 1: 43-52, 2010.
\bibitem{b17} Church, Kenneth Ward. Word2Vec. Natural Language Engineering 23, no. 1: 155-162, 2017.

\bibitem{b19} Aralikatte, Rahul, Giriprasad Sridhara, Neelamadhav Gantayat, and Senthil Mani. Fault in your stars: an analysis of android app reviews. In Proceedings of the acm india joint international conference on data science and management of data, pp. 57-66. 2018.
\bibitem{b20} Elmurngi, Elshrif Ibrahim, and Abdelouahed Gherbi. Unfair reviews detection on amazon reviews using sentiment analysis with supervised learning techniques. J. Comput. Sci. 14, no. 5: 714-726, 2018.
\bibitem{b21} Patel, Nidhi A., and Rakesh Patel. A survey on fake review detection using machine learning techniques. In 2018 4th International Conference on Computing Communication and Automation (ICCCA), pp. 1-6. IEEE, 2018.
\bibitem{b22} Kim, Jinhan, Robert Feldt, and Shin Yoo. Guiding deep learning system testing using surprise adequacy. In 2019 IEEE/ACM 41st International Conference on Software Engineering (ICSE), pp. 1039-1049. IEEE, 2019.

\bibitem{b24} Battista Biggio, Igino Corona, Davide Maiorca, Blaine Nelson, Nedim Srndi/c, Pavel Laskov, Giorgio Giacinto, and Fabio Roli. Evasion attacks against machine learning at test time. In Joint European Conference on Machine Learning and Knowledge Discovery in Databases. Springer, 387–402, 2013.
\bibitem{b25} C. Szegedy et al. Intriguing properties of neural networks. in Proc. 2nd Int. Conf. Learn. Representation, pp. 14–16, 2014.
\bibitem{b26} N. Carlini and D. A. Wagner, Adversarial examples are not easily detected: Bypassing ten detection methods/ in Proc. 10th ACMWorkshop Artif. Intell. Secur., pp. 3–14, 2017.

\bibitem{b27}Ian J. Goodfellow, J. Shlens, and C. Szegedy, Explaining and harnessing adversarial examples. in Proc. 3rd Int. Conf. Learn. Representation, pp. 7–9, 2015.
\bibitem{b28} A. Kurakin, Ian J. Goodfellow, and S. Bengio, Adversarial examples in the physical world. in Proc. 5th Int. Conf. Learn. Representation, 2017, arXiv:1607.02533.
\bibitem{b29} A. Madry, A. Makelov, L. Schmidt, D. Tsipras, and A. Vladu, Towards deep learning models resistant to adversarial attacks, in Proc. 6th Int. Conf. Learn. Representat., 2018, arXiv:1706.06083.
\bibitem{bdf} Moosavi-Dezfooli, S. M., Fawzi, A., and Frossard, P.  Deepfool: a simple and accurate method to fool deep neural networks. In Proceedings of the IEEE conference on computer vision and pattern recognition, pp. 2574-2582, 2016.

\bibitem{b30} Gao, J., Lanchantin, J., Soffa, M. L., and Qi, Y.  Black-box generation of adversarial text sequences to evade deep learning classifiers. In 2018 IEEE Security and Privacy Workshops (SPW) (pp. 50-56). IEEE. 2018, May.
\bibitem{b31} Ma, S., Liu, Y., Tao, G., Lee, W. C., and Zhang, X.  Nic: Detecting adversarial samples with neural network invariant checking. In 26th Annual Network And Distributed System Security Symposium (NDSS 2019), 2019, January.
\bibitem{b32}I.Evtimov, K.Eykholt, E.Fernandes, T.Kohno, B.Li, A.Prakash, A.Rahmati, and D. Song, Robust physical-world attacks on deep learning models, 2017. arXiv preprint arXiv:1707.08945.
\bibitem{b33} K. Pei, Y.Cao, J.Yang, and S.Jana, Deepxplore:Automated white box testing of deep learning systems. in Proceedings of the 26th Symposium on Operating Systems Principles. ACM,pp.1–18, 2017.
\bibitem{b34} Ribeiro, M. T., Wu, T., Guestrin, C., and Singh, S. (2020). Beyond accuracy: Behavioral testing of NLP models with CheckList. arXiv preprint arXiv:2005.04118.
\bibitem{b35} Ouyang, T., Isobe, Y., Sultana, S., Seo, Y., and Oiwa, Y.  Autonomous driving quality assurance with data uncertainty analysis. In 2022 International Joint Conference on Neural Networks (IJCNN) (pp. 1-7). IEEE. 2022, July.
\bibitem{b36} Ouyang, T., Marco, V. S., Isobe, Y., Asoh, H., Oiwa, Y., and Seo, Y.  Improved Surprise Adequacy Tools for Corner Case Data Description and Detection. Applied Sciences, 11(15), 6826, 2021.
\bibitem{b37} Lee, D., Yu, S., and Yu, H.  Multi-class data description for out-of-distribution detection. In Proceedings of the 26th ACM SIGKDD International Conference on Knowledge Discovery \& Data Mining, pp. 1362-1370, 2020, August.

\end{thebibliography}
\end{document}